\documentclass[journal]{IEEEtran}
\usepackage{multirow}
\usepackage{mathtools}
\usepackage{amsmath,amssymb,amsfonts}
\usepackage{graphicx}
\usepackage{textcomp}
\usepackage{xcolor}
\usepackage{subfigure} 
\usepackage[letterpaper, left=0.65in, right=0.65in, bottom=1in, top=0.70in]{geometry}

\def\BibTeX{{\rm B\kern-.05em{\sc i\kern-.025em b}\kern-.08em
    T\kern-.1667em\lower.7ex\hbox{E}\kern-.125emX}}
\usepackage{xcolor}
\usepackage[linesnumbered,ruled,vlined]{algorithm2e}
\usepackage{listings}
\usepackage[noend]{algpseudocode}

\SetCommentSty{mycommfont}
\SetKwInput{KwInput}{Input}   
\SetKwInput{KwOutput}{Output} 

\usepackage[compact]{titlesec}
\titlespacing{\section}{1.0pt}{*1.0}{*0}
\titlespacing{\subsection}{1.1pt}{*1.1}{*0}
\titlespacing{\subsubsection}{0.3pt}{*0}{*0}
\definecolor{GreenForest}{rgb}{0.09, 0.45, 0.27}
\usepackage{soul}
\usepackage{balance}
\usepackage{cite}
\usepackage{acro} 
\DeclareAcronym{BS}{short = BS,long  = Base Station ,tag    = abbrev}
\DeclareAcronym{CP-OFDM}{short = CP-OFDM ,long  = Cyclic Prefix Orthogonal Frequency Division Mltiplexing ,tag    = abbrev}
\DeclareAcronym{ISAC}{short = ISAC ,long  = Integrated Sensing And Communication ,tag    = abbrev}
\DeclareAcronym{6G}{short = 6G,long  = 6th Generation ,tag    = abbrev}
\DeclareAcronym{5G}{short = 5G,long  = 5th Generation ,tag    = abbrev}
\DeclareAcronym{4G}{short = 5G,long  = 4th Generation ,tag    = abbrev}
\DeclareAcronym{RMSE}{short = RMSE ,long  = Root Mean Squared Error ,tag    = abbrev}
\DeclareAcronym{DFT}{short = DFT,long  = Discrete Fourier Transform ,tag    = abbrev}
\DeclareAcronym{IDFT}{short = IDFT,long  = Inverse Discrete Fourier Transform ,tag    = abbrev}
\DeclareAcronym{IoT}{short = IoT,long  = Internet Of Things ,tag    = abbrev}
\DeclareAcronym{MSE}{short = MSE,long  = Mean Square Error ,tag    = abbrev}
\DeclareAcronym{SNR}{short = SNR,long  = Signal-To-Noise Ratio ,tag    = abbrev}
\DeclareAcronym{BER}{short = BER,long  = Bit Error Rate ,tag    = abbrev}
\DeclareAcronym{NMSE}{short = NMSE,long  = Normalized Mean Square Error ,tag    = abbrev}
\DeclareAcronym{AWGN}{short = AWGN,long  = Additive White Gaussian Noise,tag    = abbrev}
\DeclareAcronym{BPSK}{short = BPSK,long  = Binary Phase Shift Keying ,tag    = abbrev}
\DeclareAcronym{FMCW}{short = FMCW,long  = Frequency Modulated Continuous Wave ,tag    = abbrev}
\DeclareAcronym{JSAC}{short = JSAC,long  = Joint Sensing And Communication ,tag    = abbrev}
\DeclareAcronym{PAPR}{short = PAPR,long  = Peak To Average Power Ratio ,tag    = abbrev}
\DeclareAcronym{ML}{short = ML,long  = Maximum Likelihood ,tag    = abbrev}
\DeclareAcronym{OFDM}{short = OFDM,long  = Orthogonal Frequency Division Multiplexing ,tag    = abbrev}
\DeclareAcronym{ISI}{short = ISI,long  = Inter Symbol Interference,tag    = abbrev}
\DeclareAcronym{ICI}{short = ICI,long  = Inter Carrier Interference,tag    = abbrev}
\DeclareAcronym{LS}{short = LS,long  = Least Square,tag    = abbrev}
\DeclareAcronym{CP}{short = CP,long  = Cyclic Prefix,tag    = abbrev}
\usepackage{comment}
\usepackage{amsmath}
\usepackage{steinmetz}

\newcommand\blfootnote[1]{%
  \begingroup
  \renewcommand\thefootnote{}\footnote{#1}%
  \addtocounter{footnote}{-1}%
  \endgroup
}
\makeatletter
\newcommand{\ie}{\emph{i.e.}\@ifnextchar.{\!\@gobble}{}}
\newcommand{\eg}{\emph{e.g.}\@ifnextchar.{\!\@gobble}{}}
\newcommand{\etc}{etc\@ifnextchar.{}{.\@}}
\makeatother
\begin{document}

\bstctlcite{IEEEexample:BSTcontrol}
    \title{A Novel OFDM-FMCW Waveform for Low-Complexity Joint Sensing and Communication}
  \author{Amir Bouziane, Salah Eddine Zegrar,~\IEEEmembership{Member,~IEEE}, and H\"{u}seyin Arslan,~\IEEEmembership{Fellow,~IEEE}

\thanks{The authors are with the Department of Electrical and Electronics Engineering, Istanbul Medipol University, Istanbul, 34810, Turkey (e-mail: bouziane.amir@std.medipol.edu.tr; salah.zegrar@std.medipol.edu.tr;  huseyinarslan@medipol.edu.tr).}
\thanks{S. E. Zegrar is also with ULAK Haberleşme, Istanbul, 34906, Turkey (e-mail: salah.zegrar@ulakhaberlesme.com.tr).}
}

\IEEEpeerreviewmaketitle
\maketitle
\blfootnote{\noindent \scriptsize \copyright~2024 IEEE. Personal use of this material is permitted. Permission from IEEE must be obtained for all other uses, in any current or future media, including reprinting/republishing this material for advertising or promotional purposes, creating new collective works, for resale or redistribution to servers or lists, or reuse of any copyrighted component of this work in other works.\\
\vspace{0.2em} \\
\textbf{Citation:} A. Bouziane, S. Eddine Zegrar and H. Arslan, "A Novel OFDM-FMCW Waveform for Low-Complexity Joint Sensing and Communication," in IEEE Wireless Communications Letters, vol. 14, no. 2, pp. 425-429, Feb. 2025. \textbf{DOI:} 10.1109/LWC.2024.3506621}

\begin{abstract}
With the advent of 6th Generation (6G) wireless networks, integrating Orthogonal Frequency Division Multiplexing (OFDM) and Frequency Modulated Continuous Wave (FMCW) waveforms presents a promising solution for simultaneous communication and sensing. While OFDM offers high communication performance, it requires complex transceivers for sensing applications. Conversely, FMCW is easily implemented for sensing but provides low data rates for communication. This letter derives the mathematical relationship between OFDM and FMCW, showing how FMCW can be generated from the diagonal elements of the OFDM modulator (\ie, the Discrete Fourier Transform (DFT) matrix). We propose a novel approach that leverages the DFT matrix to seamlessly integrate these two waveforms, minimizing interference and enabling high data rate communication alongside low-complexity sensing receiver design. Extensive simulations demonstrate the effectiveness of our technique in achieving simultaneous sensing and communication.
\end{abstract}
\begin{IEEEkeywords}
6G, ISAC, OFDM, FMCW, waveform.
\end{IEEEkeywords}
\section{Introduction}
In recent years, wireless engineering research has focused on developing dual-functional base stations capable of concurrent communication and sensing tasks, aiming to eliminate the need for separate hardware. Despite hardware advancements, efficient spectrum utilization remains critical due to its scarcity and value. Modern technology has transitioned many functionalities from analog to digital platforms, offering greater waveform design flexibility. The goal is to integrate sensing and communication capabilities into a single hardware platform using a unified waveform \cite{5776640}, a key technology expected to support many applications in future wireless systems \cite{9705498}. 

\Ac{OFDM} is a premier waveform for communication due to its ability to combat frequency-selective channels and multipath propagation. However, despite its success, it cannot fulfill the diverse applications and requirements expected of \ac{6G} \cite{10382693}. One important application is sensing. Although \ac{OFDM} is applicable in sensing scenarios \cite{5599316, 7870764}, it suffers from processing overhead and target masking in the presence of multiple targets \cite{9393464}. While \ac{OFDM} relies on digital signal processing, waveforms like \ac{FMCW} can detect the range and velocity of multiple targets using simpler analog processing. However, despite its advantages in sensing, \ac{FMCW} has limitations when adapted for communication purposes \cite{7815384, 7523229, 7797659}, primarily due to its low data rates, which fails to meet the stringent requirements for high-speed communication. Integrating two waveforms to achieve both high data rates and high-resolution sensing is a viable approach. \ac{ISAC} waveforms exemplify this integration \cite{9737357}; however, this approach faces challenges, such as the waveform coexistence problem \cite{8828016} where the primary issue is potential interference between these waveforms. Various techniques, orthogonal and non-orthogonal, have been explored to address this challenge such as \cite{8782136,9348582,9771655}. For instance, in \cite{8782136}, an OFDM-FMCW waveform was generated by multiplying frequency domain \ac{FMCW} discrete data with \ac{OFDM} symbols, albeit at the cost of increased hardware overhead. Another approach, presented in \cite{9348582}, involves non-orthogonal coexistence of \ac{OFDM} and \ac{FMCW}, offering increased spectrum efficiency and flexibility but facing drawbacks such as higher \ac{BER} and reduced communication system reliability. Furthermore, researchers in \cite{9771655} utilized \ac{DFT} properties to align \ac{FMCW} signals on specific subcarriers in the frequency domain, maintaining orthogonality between the two waveforms but encountering implementation challenges in real-world scenarios.

In this paper, a novel ISAC coexistence mechanism for having OFDM and FMCW waveforms together is introduced. In particular, we extract the FMCW signal directly from the DFT matrix using diagonal elements. This method ensures minimal interference between the two waveforms and simplifies the design. Additionally, we propose a method to ensure the continuity of the \ac{FMCW} signal using \ac{CP}, which simplifies the radar receiver and enables analog processing.
The effectiveness of the proposed design is validated by simulations, demonstrating communication performance, measured by \ac{BER}, comparable to conventional \ac{OFDM} systems. Additionally, the sensing performance is evaluated using range-velocity maps, showcasing successful target detection. Further comparisons with conventional \ac{FMCW} using \ac{RMSE} simulations indicate good performance.

\textit{Notation:} Bold uppercase $\mathbf{A}$, bold lowercase $\mathbf{a}$, and unbold letters $A,a$ denote matrices, column vectors, and scalar values, respectively. $(\cdot)^{-1}$ denote the inverse operators.  $\delta(\cdot)$ denotes the Dirac-delta function.  $\mathcal{F}\{\cdot\}$ denote the Fourier transform operator. $\mathbb{C}^{{M\times N}}$ denotes the space of $M\times N$ complex-valued matrices.

\section{Proposed OFDM-FMCW Waveform}
In our proposed system, an \ac{ISAC} transceiver operates in a mono-static configuration, functioning as both the transmitter and radar receiver for sensing. It detects echoes of the transmitted signal while a separate receiver handles communication tasks. Integrating these functionalities within a single transceiver presents significant challenges, requiring a thorough understanding of the underlying signal models that support our approach.


\subsection{OFDM Signal Model}
Consider a wireless communication system where $N$ data symbols are modulated using OFDM over a total bandwidth $B$ operating at the carrier frequency $f_c$. 
The OFDM modulator distributes these symbols $\{x(m),~m=0,\dots,N-1\}$ in frequency domain. The subcarrier spacing is given as $\Delta f= B/N$ and the time duration of one OFDM symbol is $T = 1/\Delta f$. The frequency samples $x(m)$ need to be converted to time domain using the \ac{IDFT} operation. The discrete output of the $i$th \ac{OFDM} symbol modulator is expressed as 
\begin{equation}
s_i(n)=\frac{1}{\sqrt{N}} \sum_{m=0}^{N-1} x_i(m) e^{j 2 \pi m (n-L_{\text{cp}})/N},
\label{equ:ofdm}
\end{equation}
where $L_{\text{cp}}$ denotes the CP length.


\subsection{Chirp Signal Model}
Consider a linear \ac{FMCW} signal, $c(t)$, sweeping linearly across $B_c$ bandwidth during the time interval $T_c$. The time-domain representation of $c(t)$ is given by \cite{9348582}
\begin{equation}
    c(t)=e^{j\pi  \eta t^2 }, \quad 0 \leq t<T_c,
    \label{equ:chirp}
\end{equation}
where $\eta=B_c / T_c$ is the chirp rate.
Sampling \eqref{equ:chirp} at times $t=n \frac{T}{N},~ 0 \leq n<N$, produces the discrete chirp signal as follows \cite{25156940}
\begin{equation}
c(n)=e^{j\pi  \frac{\eta  T^2 n^2}{N^2} }=e^{j\pi  \frac{B_c }{T_c} \frac{T^2 n^2}{N^2} }, \quad 0 \leq n<N.
\label{equ:chirp2}
\end{equation}
Let $\gamma  = T/T_c$ be the number of chirp signals within one OFDM symbol and $B_c = B= N/T$. Then, 
\begin{equation}
    c(n) =  e^{j\pi  \gamma\frac{  n^2 }{N} } .
\label{equ:dev}
\end{equation}

\subsection{OFDM-FMCW}
The $N\times N$ \ac{IDFT} matrix $F^{-1}_N$ is defined as:
\begin{equation}
    F^{-1}_N(m,n) = \frac{1}{\sqrt{N}} e^{j2\pi \frac{mn}{N}}.
    \label{equ:dftt}
\end{equation}
From the matrix in \eqref{equ:dftt}, the ladder-like elements, denoted by the entries $m = \gamma n$, can be given by
\begin{equation}
    {C_N}(\gamma n,n) = \frac{1}{\sqrt{N}}  e^{j2\pi \gamma\frac{n^2}{N}}.
    \label{equ:diag}
\end{equation}
Let $\mathbf{c} = \mathbf{C}_N \mathbf{1}$ be the transmitted signal generated from $\mathbf{C}_N$ where $\mathbf{1}$ is the ones vector, then
\begin{equation}
    c(n) = e^{j\pi 2\gamma\frac{n^2}{N}}.
    \label{equ:cc}
\end{equation}
From \eqref{equ:dev} and \eqref{equ:cc}, it is clearly seen that the FMCW signal can be directly generated from the elements $m = \gamma n$ of IDFT matrix with a chirp rate of $\gamma' = 2\gamma$ within one OFDM symbol.

Note that for the rest of the paper we will consider the case of $\gamma = 1$ (i.e., two chirps within one OFDM symbol), which corresponds to the case where the diagonal elements of the IDFT are selected i.e., $m = \frac{2}{2} n = n$.

\begin{figure}[t]
    \centering
    \includegraphics[scale=0.7]{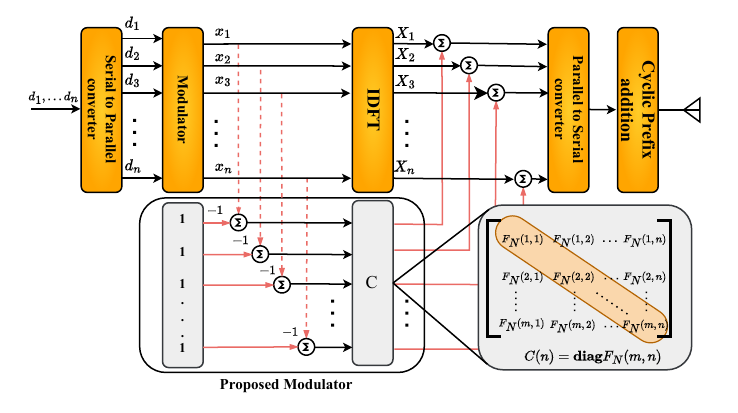}
    \caption{The proposed OFDM-FMCW block diagram.}
    \label{fig:syst_mod}
\end{figure}

\subsection{Proposed Transmitter}
Instead of the \ac{IDFT} at the transmitter, the following modulator is proposed:
    \begin{equation}
        \mathbf{S} = \boldsymbol{\Psi}^{-1} \mathbf{x},
    \end{equation}
    where $\mathbf{x}$ is our data symbols vector and  $\boldsymbol{\Psi}^{-1}$ is defined as:
    \begin{equation}
        \boldsymbol{\Psi}^{-1} = \begin{cases} {F}_N^{-1}(m,n), & \text{if } m \neq n \\ 
        {F}_N^{-1}(m,n)/{x}(n), & \text{if } m = n \end{cases}.
    \end{equation}
The transmitter equation can be written in an affine form and further simplified in terms of the conventional \ac{OFDM} system as
\begin{equation}
\begin{aligned}
    \mathbf{S} & =  (\mathbf{F}_N^{-1} - \mathbf{C}_N)\mathbf{x} + \mathbf{P}_c \mathbf{C}_N \mathbf{1}\\
    & = \underbrace{\mathbf{F}_N^{-1}\mathbf{x}}_{\text{\ac{OFDM} modulator}}+\underbrace{\mathbf{P}_c\mathbf{C}_N(\mathbf{1} - \mathbf{x})}_{\text{\ac{FMCW} signal}}.
\end{aligned}
\label{equ:mod}
\end{equation}
$\mathbf{C}_N$ is the diagonal matrix is the diagonal matrix with elements corresponding to the \ac{IDFT} diagonal. $\mathbf{P}_c$ is a design parameter representing the power ratio between the two signals. Equation \ref{equ:mod} shows the combination of the two signals in the form of an \ac{OFDM} modulator with an added \ac{FMCW} signal. This implementation is illustrated in Fig. \ref{fig:syst_mod}, where the diagonal elements are extracted from the \ac{IDFT} output, along with the data symbols, are fed into the novel modulator block.\textcolor{black}{ The generated \ac{FMCW} signal is then combined with the original \ac{OFDM} signal, converted to  time domain, and transmitted through the channel. This method is distinct from simple superposition, reducing potential interference.}


\begin{figure*}
    \centering
    \includegraphics[scale=0.412]{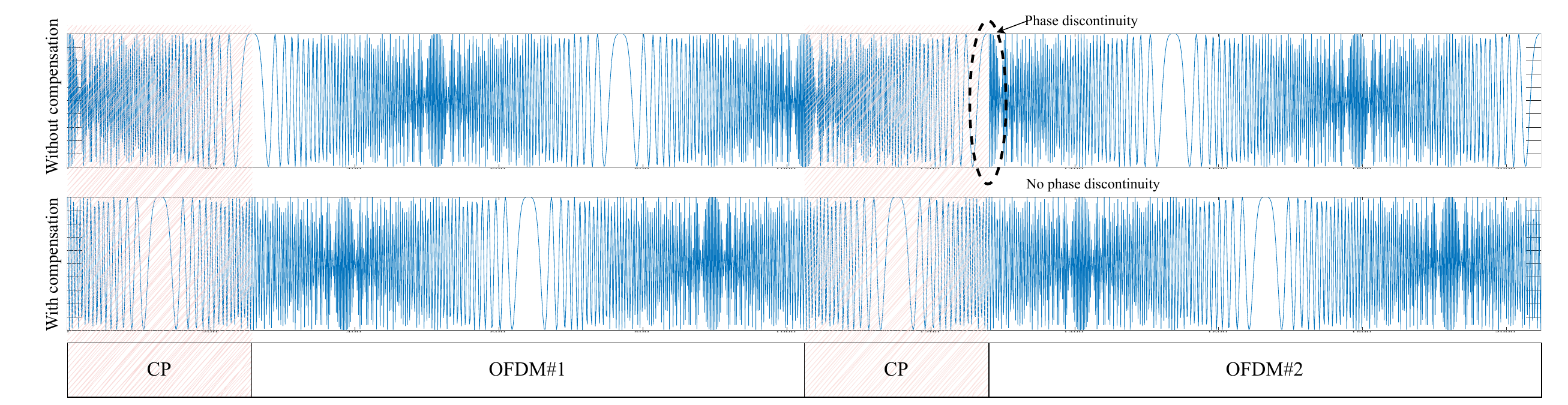}
    \caption{Appending CP in the proposed waveform.}
    \label{fig:cp}
\end{figure*}
\subsection{Cyclic Prefix}
 In conventional \ac{OFDM} systems, a \ac{CP} is appended after the parallel-to-serial conversion to ensure channel circularity and mitigate \ac{ISI}. On the other hand, from a radar perspective, the CP disrupts the continuity between the $i$-th and $(i+1)$-th chirps, causing the chirp radar to malfunction. To solve this issue of the proposed approach, a phase shift is applied to the chirp in the frequency. \textcolor{black}{This phase shift ensures an adequate time shift that compensates for the duration of the \ac{CP},  without effecting the \ac{CP}'s primary functions in \ac{OFDM} and maintaining subcarrier orthogonality.} Let $C(k) = \mathcal{F}\{c(n)\}$ be the frequency representation of the chirp, then the $i$-th chirp is given as
    \begin{equation}
        \begin{aligned}
            c_{i}(n) &= \mathcal{F}^{-1}\{C(k)\cdot e^{\frac{j2\pi (i L_{cp})k}{N}}\}\\
            & = c(n)\circledast \delta(n-i L_{cp}), \\
            & = c(n-i L_{cp}). 
        \end{aligned}
    \end{equation}
Fig. \ref{fig:cp} demonstrates how \ac{CP} is appended to $s(n)$ in detail. As depicted in Fig. \ref{fig:cp}, the chirp signal remains continuous in the time domain even with the addition of a \ac{CP}. Consequently, the de-chirping process of the received signal does not necessitate the removal of the \ac{CP} during signal processing. Instead, the signal can be directly mixed in the analog domain, as shown in Fig. \ref{fig:illustration}.

A conventional OFDM receiver can be employed for communication purposes, where the \ac{CP} is initially discarded. It is worth noting that the induced phase shift for \ac{CP} compensation does not affect communication since the location of the chirp in the time and frequency domains remains fixed.


\section{Channel}
\begin{figure}[t]
    \centering
    \includegraphics[scale=0.75]{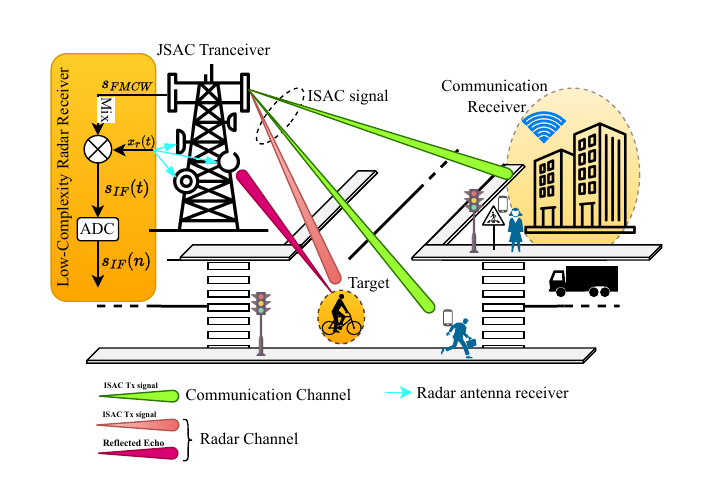}
    \caption{Illustration of the ISAC system.}
    \label{fig:illustration}
\end{figure}

An \ac{ISAC} \ac{BS} is transmitting a uniform \ac{ISAC} waveform in the environment. Multiple users are present with probable objects to detect. The \ac{BS} transmits data to $K$ single-antenna communication users while simultaneously extracting environmental information from reflected echoes as shown in Fig. \ref{fig:illustration}. A communication channel is assumed for the users, and a radar channel is designated for the objects due to the processing of the radar signal that occurs in the \ac{BS}.
\subsection{Communication Channel}
The impulse response of a wireless channel can be expressed as \cite{4607239}:
\begin{equation}
h(t) =  \sum_{l=1}^{L} \sigma_l \delta(t-\tau_l),
\end{equation}
where $\tau_l$ and $\sigma_l$ are the delay and complex amplitude of the $l$-th path, respectively. The received signal can then be expressed as
\begin{equation}
y(t) =  \sum_{l=1}^{L} h_l s \delta(t-\tau_l) + w(t),
\end{equation}
where $h_l$ represents the channel taps, and $w(t)$ is the \ac{AWGN} noise with $\mathcal{N}(\mu,,\sigma^{2})$ at the receiver. After down-converting the received signal and sampling at the rate of $f_s$, the discrete-time domain signal for one \ac{OFDM} symbol time interval can be written as:
\begin{equation}
y(k) = \sum_{l=1}^{L} h_l s(k-\tau_l) + w(k),
\end{equation}
where $k=0,1,...,\lambda-1$, with $\lambda = T_s \times f_s$. After removing the \ac{CP} and performing \ac{DFT} operations, we obtain $y \in \mathbb{C}^{N \times 1}$, the frequency domain vector.\textcolor{black}{ Diagonal elements representing our chirp signal are used as pilots instead of the conventional \ac{OFDM} pilot allocation, which confines interference to data symbols, allowing for accurate channel estimation and improved \ac{BER} performance compared to direct superposition methods.} The \ac{LS} estimator is used \cite{Tan2011}.

\begin{equation}
    \begin{aligned}
        \mathbf{Y} & =  \mathbf{F}_N \mathbf H  \mathbf{F}_N^{-1}  \mathbf{x} +\mathbf{F}_N \mathbf H  \mathbf{P_c}  \mathbf{C}_N^*(\mathbf{1} - \mathbf{x}) + \mathbf{W}, \\
         & =   
        \mathbf{F}_N \mathbf H  \mathbf{F}_N^{-1}  \mathbf{x} +  \underbrace{\mathbf{P_c}(\underbrace{\mathbf{F}_N \mathbf H  \mathbf{C}_N^* \mathbf{1}}_{\text {Known interference}}-  \underbrace{\mathbf{F}_N \mathbf H \mathbf{x} + \mathbf{W}}_{\text {G}})}_{\text{Radar interference}}.
        \label{equ:12}
   \end{aligned} 
\end{equation} Here, $\mathbf{H}$ is the circular channel matrix, and we consider a time-invariant channel model. Radar interference presents the contribution of both the Known interference which can be removed at the receiver and $G$ which is considered noise and can be filtered out. We have $\text{SNR}_s = P_c/\sigma^2$.


\subsection{Chirp Processing}

For our scenario, our transmitter also serves as our radar receiver or can be low-end devices intended for sensing, easily implementable in existing \ac{IoT} devices. Let us consider a $R$-tap doubly-selective radar channel modeling $R$ targets in the environment with $r_p$ range and $v_p$ velocity as follows
\begin{equation}
g(\tau,\nu) = \sum_{p=0}^{R-1} g_p\delta(t-\tau_p) e^{j2\pi\nu_p(\tau-\tau_p)},
\end{equation}
where $g_p$ represents the complex channel gain coefficient corresponding to the $p$-th path, $\tau_p={2r_p}/c$, and $\nu_p={2v_pf_c}/c$ denote the round trip delay and Doppler, respectively; $c$ is the speed of light. The output of the radar channel is given as
\begin{equation}
r(t)=\sum_{p=0}^{R-1} g_p y(t-\tau_p)e^{j2\pi\nu_p(t-\tau_p)}.
\end{equation}

\begin{figure}[t]
    \centering
    \includegraphics[scale=0.7]{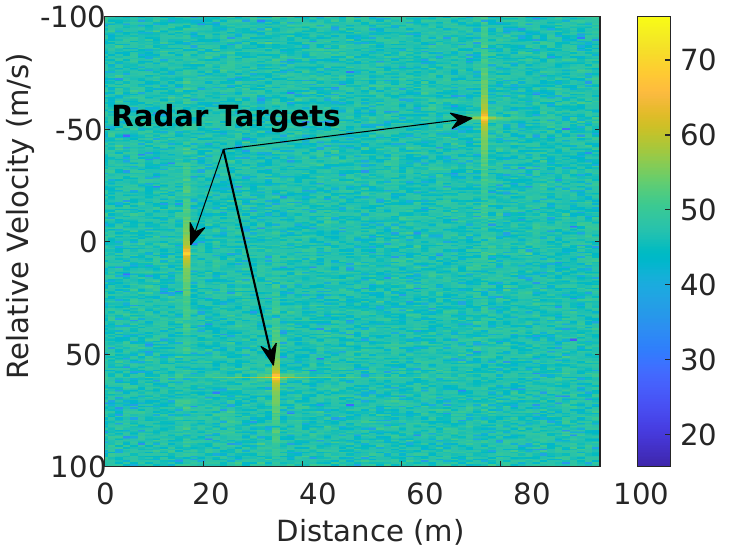}
    \caption{Range Doppler map}
    \label{fig:RDM}
\end{figure}

The signal reflected from this channel mixes with the transmitted chirp signal $c^*(t)$ to create a low-frequency beat signal. Our mixed signal can be written as
\begin{equation}
r_m(t) =\sum_{p=0}^{k-1} \tilde{h}_p e^{j2\pi(\eta\tau_p + \nu p)t} + a(t),
\label{equ:15}
\end{equation}
where
\begin{equation}
a(t) = w(t)+\sum_{p=0}^{k-1} h_p y(t - \tau_p) e^{j2\pi\nu_p(t-\tau_p)} \cdot c^*(t).  
\end{equation}
$a(t)$ is the communication or data interference on the \ac{FMCW} signal, which is the contribution of the communication signal and additive white noise. It will be represented in the simulations by $\text{SNR}_c = P_x/\sigma^2$ where $P_x$ is the power of the data of our communication signal.  $\tilde{h}_p$ is the complex amplitude of the $p$-th path; simplified terms are not time-variant and therefore we simplify to obtain \eqref{equ:15}.

When the target is stationary, the beat frequency is determined only by the time delay and frequency change rate of the waveform. However, when the target is in motion, the beat frequency incorporates the Doppler shift in addition to the delay. The sampling of \eqref{equ:15} gives us the beat frequency $f_b=\eta \tau_p + \nu_p$. Hence, the beat frequency $f_b$ can be represented as
\begin{equation}
f_b = \frac{2r_p B}{cT} + \nu_p.
\label{equ:beatf}
\end{equation}

The count of beat frequencies corresponds to the count of targets. In equation \eqref{equ:beatf}, the range $r_p$ and velocity $v_p$ of the targets are coupled. Hence, to separate these variables, it is necessary to transmit multiple chirp sweeps. Subsequently, a 2-dimensional \ac{DFT} is employed on the beat frequencies to construct a range-velocity map. The resolution of the targets in terms of distance ($\Delta p$) and velocity ($\Delta v$) is determined accordingly
\begin{equation}
\Delta p = \frac{c}{2B}, \quad \Delta v = \frac{\lambda}{2T}.
\label{equ:velorang}
\end{equation}
\section{Simulation Results}
In this section, we evaluate the communication and sensing performance of the proposed waveform. For sensing, we present the range-velocity plot and evaluate its effectiveness. For communication, we assess the Bit Error Rate (\ac{BER}) and compare it to conventional \ac{OFDM}.\textcolor{black}{The simulation parameters were set as follows: $f_c$ was chosen to be $77$ GHz, $B$ of $200$ MHz was utilized. The choice of $f_c$ and $B$ is to achieve higher resolution in the range-doppler map while demonstrating the feasibility of integrating satisfactory sensing capabilities within an OFDM-based ISAC system.} Each symbol employed $1024$ sub-carriers, with a subcarrier spacing of $\frac{B}{M}$ MHz, where $M$ is the number of sub-carriers per symbol. The simulation utilized $100$ \ac{OFDM} symbols. Targets were simulated within a range uniformly distributed between $10$ and $80$ meters for their distances ($r_p$) and within a range uniformly distributed between $-70$ and $70$ meters per second for their velocities ($\nu_p$).

Fig. \ref{fig:RDM} illustrates the range-velocity map for the proposed scenario with three randomly moving targets. The 2-dimensional \ac{DFT} yields the beat frequency, coupled with Doppler frequency $\nu_p$. This range transformation isolates the Doppler shift associated with a specific target range, making targets and their velocities clearly distinguishable.

\begin{figure}[t!]
    \centering
   \subfigure[]{\includegraphics[width=0.35\textwidth]{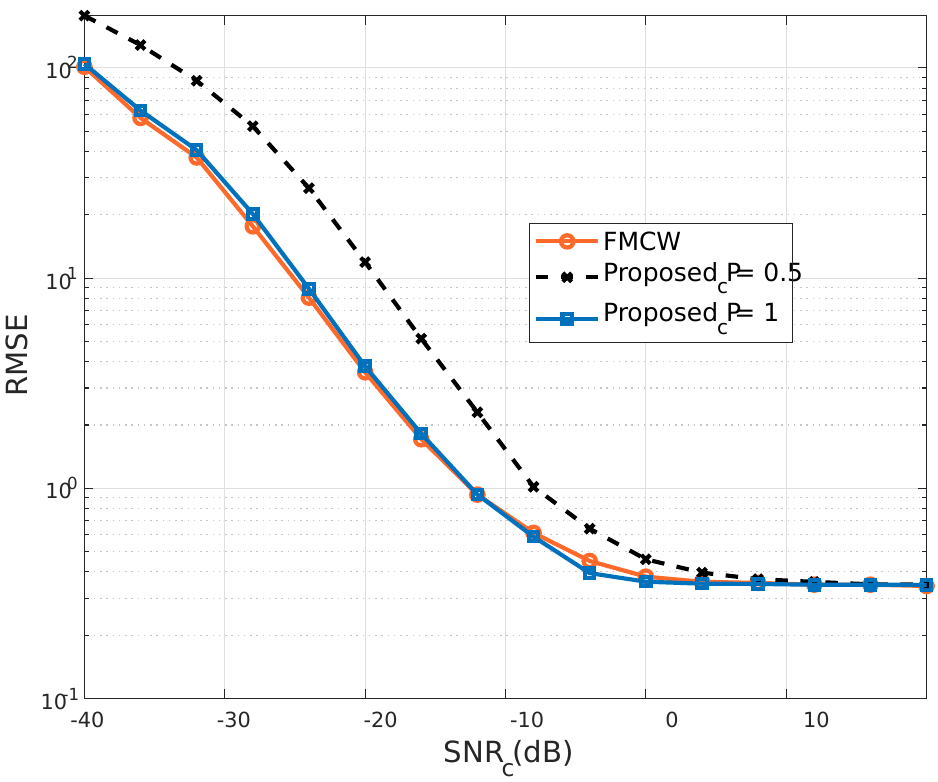}\label{fig:sub-second}} 
    \subfigure[]{\includegraphics[width=0.35\textwidth]{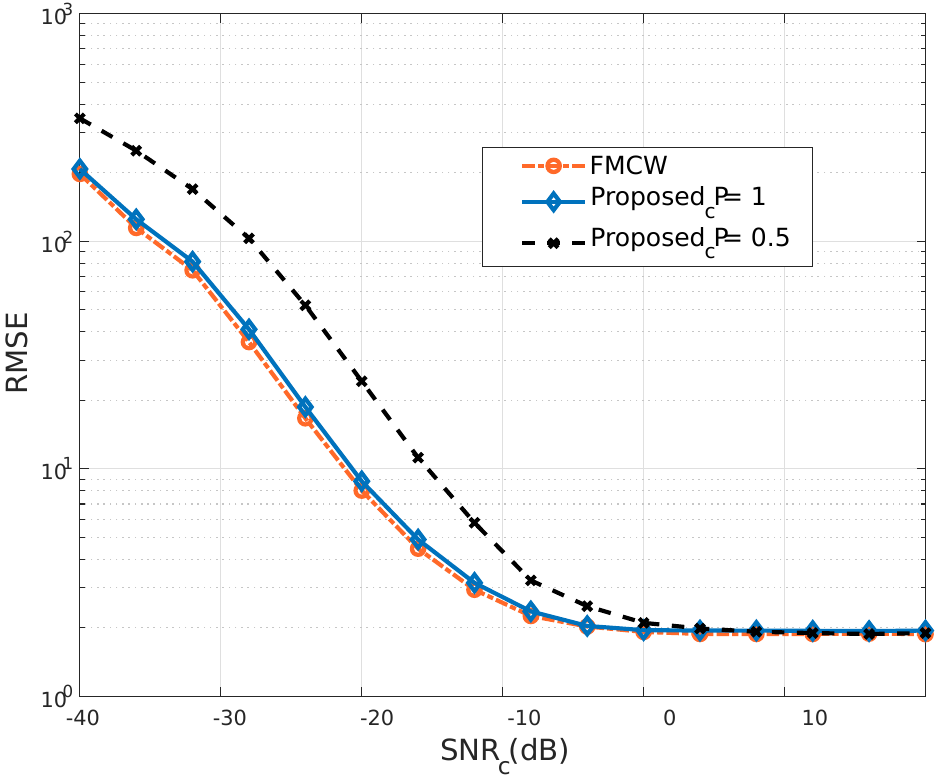}\label{fig:sub-third}}
    \caption{RMSE performance  of: (a) range, and (b) velocity vs SNR. }
    \label{fig:sens_perf}
\end{figure}

Fig. \ref{fig:sub-second} and \ref{fig:sub-third} compare the proposed waveform's Root Mean Square Error (\ac{RMSE}) performance for both range and velocity with that of conventional \ac{FMCW}. The simulation assumed a single target with uniformly distributed range and velocity. Notably, the proposed \ac{ISAC} waveform achieves comparable performance to FMCW across varying transmit powers. Additionally, the \ac{RMSE} improves for both range and velocity with increasing chirp power $P_c$.

\begin{figure}[t]
    \centering
    \includegraphics[scale=0.43]{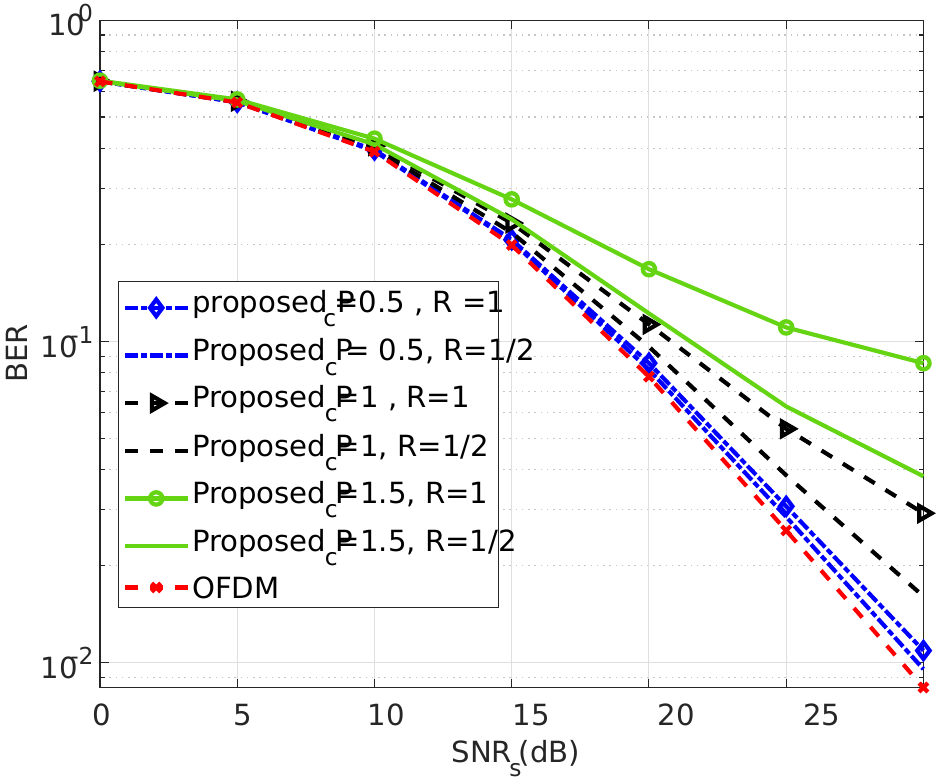}
    \caption{BER vs SNR analysis}
    \label{fig:performance}
\end{figure}

Fig. \ref{fig:performance} showcases the \ac{BER} performance of \ac{OFDM} and the proposed design for various configurations. Tested with different $P_c$ values ($P_c \in (0.5,1,1.5)$) with different pilot ratios $R$ as a parameter influencing this performance. We assumed perfect channel estimation for both designs during the evaluation. The \ac{BER} performance for the proposed design is presented across different values of the \ac{FMCW} parameter $Pc$ and $R$. As expected, under identical parameters, the proposed design with a pilot ratio of $1$ exhibits \ac{BER} performance comparable to conventional \ac{OFDM}. However, by varying $Pc$ and employing lower pilot ratios $R < 1$, a degradation in \ac{BER} performance becomes evident. This is attributed to the increasing interference power introduced by the \ac{FMCW} samples within our \ac{IDFT} matrix.

\section{Conclusion}
In this letter, we introduce a novel waveform design for an \ac{ISAC} system, integrating two initially separate waveforms, \ac{OFDM} and \ac{FMCW}, into one unified waveform. Through the extraction and reintroduction of \ac{FMCW} via a novel modulator and \ac{CP} appending mechanism, seamless integration is achieved. The proposed solution is adaptable to existing systems and offers straightforward implementation in radar receivers. Simulations attest to the effectiveness and reliability of this approach in \ac{ISAC} systems. 
Furthermore, our study reveals a trade-off between communication and sensing performance in the overall system. Depending on the specific requirements for both functionalities, adjustments to parameters such as $P_c$ and pilot ratio can optimize system performance.

\IEEEpeerreviewmaketitle
\bibliographystyle{IEEEtran}
\bibliography{Bibliography}

@ARTICLE{Tan2011,
  author={Tan, J. and others},
  journal={Sci. China Inf. Sci.},
  title={An optimal adaptive filtering algorithm with a polynomial prediction model},
  year={2011},
  volume={54},
  pages={153–162},
  doi={10.1007/s11432-010-4141-3}
}

@ARTICLE{10382693,
  author={Solaija and others},
  journal={IEEE Veh. Technol. Mag}, 
  title={Orthogonal Frequency Division Multiplexing: {The} Way Forward for {6G} Physical Layer Design?}, 
  year={2024},
  volume={19},
  number={1},
  pages={45-54},
  doi={10.1109/MVT.2023.3344432}}

@ARTICLE{4607239,
  author={Hwang and others},
  journal={IEEE Trans. Veh. Technol}, 
  title={{OFDM} and Its Wireless Applications: A Survey}, 
  year={2009},
  volume={58},
  number={4},
  pages={1673-1694},
  doi={10.1109/TVT.2008.2004555}}

@ARTICLE{8828016,
  author={Zheng and others},
  journal={IEEE Signal Process. Mag}, 
  title={Radar and Communication Coexistence: An Overview: A Review of Recent Methods}, 
  year={2019},
  volume={36},
  number={5},
  pages={85-99},
  doi={10.1109/MSP.2019.2907329}}

@ARTICLE{9737357,
  author={Liu, Fan and others},
  journal={IEEE J. Sel. Areas Commun.}, 
  title={Integrated Sensing and Communications: Toward Dual-Functional Wireless Networks for 6{G} and Beyond}, 
  year={2022},
  volume={40},
  number={6},
  pages={1728-1767},
  doi={10.1109/JSAC.2022.3156632}}

@ARTICLE{9705498,
  author={Liu, An and others},
  journal={IEEE Commun. Surveys Tuts.}, 
  title={A Survey on Fundamental Limits of Integrated Sensing and Communication}, 
  year={2022},
  volume={24},
  number={2},
  pages={994-1034},
  doi={10.1109/COMST.2022.3149272}}

@article{25156940,
  title={A Novel {OTFS-Chirp} Waveform for Low-Complexity Multi-User Joint Sensing and Communication},
  author={Zegrar and others},
  journal={Authorea Preprints},
  year={2024},
  publisher={Authorea}
}

@ARTICLE{5776640,
  author={Sturm and others},
  journal={IEEE}, 
  title={Waveform Design and Signal Processing Aspects for Fusion of Wireless Communications and Radar Sensing}, 
  year={2011},
  volume={99},
  number={7},
  pages={1236-1259},
  doi={10.1109/JPROC.2011.2131110}}

@ARTICLE{7870764,
  author={Patole and others},
  journal={IEEE Signal Processing Mag}, 
  title={Automotive radars: A review of signal processing techniques}, 
  year={2017},
  volume={34},
  number={2},
  pages={22-35},
  doi={10.1109/MSP.2016.2628914}}

@ARTICLE{7815384,
  author={Raza and others},
  journal={IEEE Communications Surveys \& Tutorials}, 
  title={Low Power Wide Area Networks: An Overview}, 
  year={2017},
  volume={19},
  number={2},
  pages={855-873},
  doi={10.1109/COMST.2017.2652320}}

@ARTICLE{7523229,
  author={Ouyang and others},
  journal={IEEE Trans. Commun}, 
  title={Orthogonal Chirp Division Multiplexing}, 
  year={2016},
  volume={64},
  number={9},
  pages={3946-3957},
  doi={10.1109/TCOMM.2016.2594792}}

@INPROCEEDINGS{7797659,
  author={Reynders and others},
  booktitle={2016 Symposium on Communications and Vehicular Technologies (SCVT)}, 
  title={Chirp spread spectrum as a modulation technique for long range communication}, 
  year={2016},
  volume={},
  number={},
  pages={1-5},
  doi={10.1109/SCVT.2016.7797659}}

@ARTICLE{8782136,
  author={Jia and others},
  journal={IEEE Syst. J}, 
  title={Integrated Communication and Localization System With {OFDM}-{Chirp} Waveform}, 
  year={2020},
  volume={14},
  number={2},
  pages={2464-2472},
  doi={10.1109/JSYST.2019.2929254}}

@INPROCEEDINGS{9348582,
  author={Mert Şahin and others},
  booktitle={2020 IEEE 92nd Vehicular Technology Conference (VTC2020-Fall)}, 
  title={Multi-functional Coexistence of Radar-Sensing and Communication Waveforms}, 
  year={2020},
  volume={},
  number={},
  pages={1-5},
  doi={10.1109/VTC2020-Fall49728.2020.9348582}}

@INPROCEEDINGS{9771655,
  author={Memisoglu and others},
  booktitle={2022 IEEE Wireless Communications and Networking Conference (WCNC)}, 
  title={Orthogonal Coexistence of Overlapped Radar and Communication Waveforms}, 
  year={2022},
  volume={},
  number={},
  pages={2190-2195},
  doi={10.1109/WCNC51071.2022.9771655}}

@ARTICLE{9393464,
  author={Luong and others},
  journal={IEEE Communications Surveys \& Tutorials}, 
  title={Radio Resource Management in Joint Radar and Communication: A Comprehensive Survey}, 
  year={2021},
  volume={23},
  number={2},
  pages={780-814},
  doi={10.1109/COMST.2021.3070399}}

@ARTICLE{5599316,
  author={Sen and others},
  journal={IEEE Trans. Signal Process}, 
  title={Adaptive {OFDM} Radar for Target Detection in Multipath Scenarios}, 
  year={2011},
  volume={59},
  number={1},
  pages={78-90},
  doi={10.1109/TSP.2010.2086448}}

\vfill

\end{document}